\documentclass[fleqn,usenatbib]{mnras}
\usepackage{newtxtext,newtxmath}

\usepackage[T1]{fontenc}

\DeclareRobustCommand{\VAN}[3]{#2}
\let\VANthebibliography\thebibliography
\def\thebibliography{\DeclareRobustCommand{\VAN}[3]{##3}\VANthebibliography}

\usepackage{graphicx}	
\usepackage{amsmath}
\usepackage{amsfonts} 
\usepackage{makecell}
\usepackage{array}
\usepackage{multicol}
\usepackage{mathtools}
\usepackage{url}
\usepackage{siunitx} 
\usepackage{xspace}
\usepackage{anyfontsize}
\usepackage{float}
\usepackage{scalerel}
\usepackage{dirtytalk}
\newcommand\HII{H\protect\scaleto{$II$}{1.2ex}}
\usepackage{tikz}
\graphicspath{{figures/}}

\newcommand{\cmsq}{\,cm$^{-2}$\xspace}     
\newcommand{\Msun}{\,M$_\odot$\xspace}     
\newcommand{\ergs}{\,erg\,s$^{-1}$\xspace} 
\newcommand{\kms}{\,km\,s$^{-1}$\xspace}   

\title[The AGN SED of NGC 1068]{A consistent radio to sub-mm pc-scale study of the nucleus of NGC 1068}

\author[I. M. Mutie et al.]
{Isaac M. Mutie,$^{1,2}$ \thanks{E-mail: mumoisaac@gmail.com}
Santiago del Palacio,$^{3}$
Robert J. Beswick,$^{2}$
David Williams-Baldwin,$^{2}$
\newauthor
Jack F. Gallimore,$^{4}$
John S. Gallagher,$^{5,6}$
Susanne E. Aalto$^{3}$
and Paul O. Baki,$^{1}$
\\
$^{1}$Department of Astronomy and Space Science, Technical University of Kenya, P.O Box 52428 $-$ 00200, Nairobi, Kenya\\
$^{2}$Jodrell Bank Centre for Astrophysics, School of Physics and Astronomy, The University of Manchester, Manchester M13 9PL, UK\\
$^{3}$Department of Space, Earth and Environment, Chalmers University of Technology, 412 96 Göteborg, Sweden\\
$^{4}$Department of Physics and Astronomy, Bucknell University, Lewisburg, PA 17837, USA\\
$^{5}$Department of Physics and Astronomy, Macalester College, 1600 Grand Ave, St. Paul, MN 55105 USA\\
$^{6}$Department of Astronomy, University of Wisconsin-Madison, 475 North Charter St., Madison WI 53706, USA}

\date{Accepted XXX. Received YYY; in original form ZZZ}

\pubyear{\the\year{}}

\begin{document}
\label{firstpage}
\pagerange{\pageref{firstpage}--\pageref{lastpage}}
\maketitle

\begin{abstract}
The origin of radio emission in radio-quiet (RQ) AGN remains a long-standing mystery. We present a detailed study of the cm to sub-mm emission from the nucleus of the nearby prototypical RQ Seyfert 2 galaxy, NGC~1068. We analyse observations between 4.5--706~GHz using $e$-MERLIN, VLA and ALMA. We restricted all data used for imaging to a matching $uv-$range of 15$-$3300~k$\lambda$, to ensure that all data sampled the same spatial scales. All images were restored with a $\sim$~0.06$''$ beam. To derive the spectral energy distribution (SED), we fit synchrotron, free--free, dust and coronal component models to these data. We report that the sub-mm excess between $\sim$~200--700~GHz is consistent with synchrotron emission from a compact and optically thick corona with a radius $R_\mathrm{c}\approx 70\pm5 \,R_\mathrm{g}$, a fraction of $\sim$\,$10\pm2$\% of the energy density in the form of non-thermal electrons, and a magnetic field strength $B\approx 148$~G. The luminosity of the corona is roughly consistent with---though higher than---the expected from mm--X-ray correlations derived in recent studies of RQ AGN. This difference is likely due to the corona SED peaking at ($\approx$550~GHz). Between 10 and $\sim$~200~GHz, the SED is dominated by free--free emission. High angular resolution observations at frequencies below 5~GHz are needed to constrain a putative optically thin synchrotron component and the amount of free--free absorption.

\end{abstract}



\section{Introduction}

NGC~1068 is a nearby prototypical radio-quiet (RQ) Seyfert 2 galaxy, redshift$(z)=0.00379$, (optical velocity)$V_{optical} = 1148$~km\,s$^{-1}$, spatial scale $=$ 72 pc/arcsec, \citep{Tully_1994, capetti_HST_1995,Bland-Haw_etal_1997}. The accretion disk of NGC~1068 and the broad-line region (BLR) are obscured by a thick dusty, molecular torus detected only by spectropolarimetry \citep{Antonucci&Miller1985, Mason_2006, ramos-2017Nature, Imanishi_2018}. NGC~1068 has become the archetype for obscuration-based unifying models \citep{Antonucci_1993}. The X-ray emission measured by \textit{NuSTAR}, \textit{XMM-Newton} and \textit{Chandra} is largely attenuated and variable, indicating that the obscuring medium is Compton-thick and clumpy \citep{young-2001-x-ray, capetti_2013A, Mason_2006, marinucci-2016MNRAS}.

A flat spectrum radio source, dubbed S1, marks the location of a supermassive black hole (SMBH) on 20-mas ($\sim$\,1.4~pc) scales \citep{Gallimore_1996GTrue,gallimore-2024}. Very Long Baseline Array (VLBA) observations resolve the continuum source into an elongated, pc-scale structure oriented nearly at right angles to the kpc-scale radio jet (which is nearly north-south). The morphology of S1 has been interpreted as a plasma torus \citep{Gallimore2004_Parsec} located inside the obscuring molecular torus resolved by ALMA \citep{imanishi-2016-gas, Impellizzeri_2019, garcia-2019-core}.

At radio--sub-mm wavelengths, S1 (see Fig.~\ref{fig:combined}) has a peculiar continuum spectrum compared to the central radio sources in other Seyfert nuclei of comparable AGN luminosity. Typically, radio emission in Seyfert nuclei is attributed to steep-spectrum synchrotron emission \citep[e.g.][]{ulvestand-wilson-seyferts-1989}; however, the cm-wave spectrum of S1 is flat, steep, or inverted depending on the angular resolution of the interferometric observations used to derive the flux density. In other words, the radio emission of S1 has a complicated morphology and is resolved on all scales, and comparing data sets with very widely varying $uv-$coverage and angular resolution has so far only led to conflicting results \citep[e.g.][]{ulvestand-wilson-seyferts-1989,Gallimore2004_Parsec,Krips_2006,Inoue_2020,baskin-2021,michiyama-2023}. As a result, it has been challenging to construct a reliable spectral energy distribution (SED) for the nuclear radio source S1, and understanding the emission mechanism(s) responsible for the radio continuum has remained a long-standing problem.

Despite the challenges of measuring the radio--sub-mm wavelength SED of the AGN in NGC 1068, there is evidence that thermal free--free (FF) emission contributes significantly to the cm--sub-mm spectrum. Integrated over the source (component S1), the cm-wave spectrum is flat between $\sim$\,5--$\sim$\,200~GHz \citep{muxlow-1996,gallimore_radios_sub_arcsec,Inoue_2020,michiyama-2023}. The estimated brightness temperature at 5~GHz from VLBA observations is $T_\mathrm{B}\,=\,4\,\times\,10^6$~K, which is too low for synchrotron self-absorption (SSA) to be important in shaping the SED \citep{Gallimore2004_Parsec}. The radio morphology of S1 roughly matches the warmest parts of the infrared (IR) emitting dusty torus and the distribution of water megamasers \citep{Gamez_Rosas_2022}. Thermal FF emission would imply the presence of a 10$^{6}$~K plasma intimately connected with dense molecular clouds traced by the masers and dust continuum \citep{Gallimore_nuclear_watermaser_2001,Gallimore2004_Parsec,gallimore-masers-2023}. This picture postulates a highly efficient plasma heating mechanism; the FF emission amounts to $\sim$\,20\% of the AGN bolometric luminosity \citep[scaling from][]{Gallimore2004_Parsec}, and for this reason, several authors have proposed alternative explanations \citep[e.g.][]{Inoue_2020, baskin-2021}. The only way to settle this decades-long problem and to reveal the true nature of S1 is to construct a $bona-fide$ SED of the AGN of NGC 1068 to measure and separate the emission processes on parsec-scales. To accomplish this goal, we need to measure the SED with sufficiently high, matching angular resolution and $uv$-coverage.

Identifying the origin of the S1 radio continuum source of NGC~1068 is fundamental to our understanding of the obscuring region surrounding the AGN. The nature of S1 must be intimately tied to the heating, dynamics, and resulting geometry of the molecular torus, all of which are central to understanding AGN unified schemes. In particular, synchrotron emission from an optically thick corona can be expected to produce a \say{bump} in the mm/sub-mm SED \citep[e.g.][]{inoue-2014, behar_2015, ricci-2023, Shablovinskaya-2024}; such a spectral feature has already been suggested in NGC~1068 \citep{Inoue_2020, michiyama-2023}. Proper modelling of this component can yield reliable estimates of the coronal size, non-thermal particle content, and magnetic field strength. These characterisations can then be used to improve the recent proposition that neutrinos detected in NGC~1068 \citep{icecube_2020} originate in the X-ray corona \citep{Inoue_2020, Eichmann-2022, Padovani_2024}, in the vicinity of the $(1.2$--$1.7)\times 10^{7}\,\mathrm{M}_{\odot}$ SMBH hosted by NGC~1068 \citep{maser_Greenhill_1997, GRAVITY2020,gallimore-2024}. Recent studies have proposed that neutrinos can be produced within the X-ray corona because accelerated electrons and protons in the coronae can generate gamma-rays and neutrinos through inverse Compton scattering, $pp$ and $p\gamma$ interactions \citep{Inoue_2018,Inoue_2019,Inoue_2020}.

In this paper we use data from the $enhanced$-Multi Element Linked Interferometer Network ($e$-MERLIN), the Karl G. Jansky Very Large Array (VLA), and the Atacama Large Millimeter/sub-millimeter Array (ALMA) to obtain a complete radio--sub-mm SED with data points between 4.5 and 706\,GHz, with matched angular resolution ($\sim$~0.06$''$) and $uv-$range (15$-$3300~k$\lambda$). In addition, we include complementary data at 1.4 GHz from the $e$-MERLIN to further constrain the SEDs at lower frequencies.

The structure of this paper is such that: in Sect.~\ref{data-reduction} we present the calibration, reduction and imaging of the data. In Sect.~\ref{sec:sed-fitting} we discuss the SED fitting model. In Sect.~\ref{results&discussion} we discuss the results obtained. In Sect.~\ref{summary}, we conclude from our findings and discuss future work.

\section{Observations, data reduction and imaging}\label{data-reduction}

The $e$-MERLIN data were observed on 2022 March 24, 25, 28, April 22, May 20 and 22, under project code: CY13006. A total of 24 spectral windows (spws) were observed, each with a bandwidth of 128 MHz. Each observing epoch covered 4 spws spanning 512 MHz, from 4.5 to 7.5~GHz and were centred at 4.8, 5.3, 5.8, 6.3, 6.8 and 7.3 GHz respectively. Observations for the 3$^\mathrm{rd}$ epoch (28$^\mathrm{th}$) March) were unsuccessful and hence were not included in the subsequent analysis. Observations were interspersed between the phase calibrator (J0239--0234) and the target source, NGC~1068. Target scans were 6 minutes long, while phase calibrator scans were 2 minutes long. 3C286 and OQ208 were observed for flux density and bandpass calibration respectively. The total time on target was $\sim$\,42~hrs across all epochs. Data were calibrated using the eMCP \verb|CASA| pipeline version: v1.1.19 with \verb|CASA| version: 5.8.0 \citep{casa_McMulllin2007,eMCP-javier}. Data from all epochs were self-calibrated and then divided into the respective 512 MHz bands for imaging to enhance the frequency sampling of the SED. 

The $e$-MERLIN data were observed between 1.25--1.76~GHz, with central frequency at 1.5~GHz. Data were observed in three epochs in 2018 January 9, 10 and 12, under project ID: CY6216, with 8 spws of 64 MHz width. The same calibration sources were used as in the 4.5--7.5~GHz observations above. The total observation time per epoch was 10 h 45 min, with scans interleaved between the target source and phase calibrator. Target scans were 7 min long, while phase calibrator scans were 3 min long. The total time on source target was $\sim$\,21 hrs.

\begin{table*}
    \centering
    \caption{Observations and data used.}
    \label{tab:obs}
    \begin {tabular}{llcccr }
    \hline
        Frequency (GHz)& Obs' date& Interferometer&Project ID& Time (min) & Comments  \\
        \hline\hline
        1.25--1.76 &2018 Jan  9--12 &$e$-MERLIN & CY6216 & 1260& This work \\
        4.5--7.5 &2022 Mar--May & $e$-MERLIN& CY13006 & 3060& This work \\
       8--12 & 2021 Jan 20,26& VLA &38677764 &48 & Archival \citep{mutie-2024}\\
        15  & <1985 Nov 5 &VLA & -- &$\sim 195$ & Archival \citep{wilson_Ulvestad_EVN_1987}\\
      18.5--23.5& 2015 Jun--Jul &VLA &15A-345  &180 & Archival \citep{mutie-2024}\\
       93--106& 2019 Jun 9 &ALMA & 2018.A.00038.S &26 & Archival \citep{maeda-2023} \\  
      241--257& 2021 Sep 23 &ALMA &2017.1.01666.S &60 & Archival \citep{Impellizzeri_2019} \\
         345--356& 2017 Sep 8&ALMA & 2016.1.00232.S &50 & Archival \citep{garcia-2019-core}\\
        477--489& 2022 Sep 8& ALMA& 2021.1.00279.S & 26& Archival (unpublished) (PI: Dieu Nguyen)\\
        688--706& 2015 Sep 25 &ALMA & 2013.1.00055.S & 34 & Archival \citep{GarciaTorus2016}\\
        \hline
    \end{tabular}
\end{table*}

All the ALMA data used in this work (see Table~\ref{tab:obs}) were requested through the ALMA help desk. They used \verb|CASA| pipeline version 6.5.4.9 to calibrate these data, then split the target source, averaged the data in time and frequency to minimise on storage and computing time, while avoiding time and bandwidth smearing. These data were downloaded and inspected while flagging any obvious bad data (including any spectral lines). Clean data was self-calibrated and imaged per band. The details of reduction and imaging of the VLA data, 8$-$12 GHz and 18.5$-$23.5 GHz respectively, used in this work as summarised in Table~\ref{tab:obs} have been described in \cite{mutie-2024}. The native angular resolution of VLA 8$-$12 GHz data is $\sim$~0.15$''$ with robust~$-$2 of Briggs weighting. During the imaging and deconvolution we applied a $uv$-range of 15~k$\lambda$-$3300$~k$\lambda$ and restoring beam was set to $0.06''$ (circular).

\subsubsection*{Imaging}

Imaging procedures for all $e$-MERLIN 4.5--7.5~GHz and all VLA data (4--12~GHz and 18--23~GHz) are detailed in \citet{mutie-2024}. The $e$-MERLIN 1.25--1.76~GHz data was imaged in \verb|CASA| task \verb|TCLEAN|. In all ALMA datasets, imaging was performed in \verb|CASA| with task \verb|TCLEAN|. A multi-frequency multi-term synthesis (mtmfs) deconvolver was used with a Taylor polynomial of 2. This was standardised for all data in this work. For the $e$-MERLIN 4.5--7.5~GHz, all the VLA and ALMA data, the restoring beam and $uv-$range were matched at $\sim$~0.06$''$ and $\sim$~15$-$3300~k$\lambda$ (determined by the range of overlapping spatial scales available within these data sets), respectively during imaging stage in \verb|CASA| task \verb|TCLEAN|. This ensures that all data sets used are sensitive to the same range of source spatial scales, and that source flux densities can be measured in the same physical regions.
\begin{figure}
    \includegraphics[width=\linewidth, trim=0.3cm 0.325cm 0.3cm 0.31cm, clip]{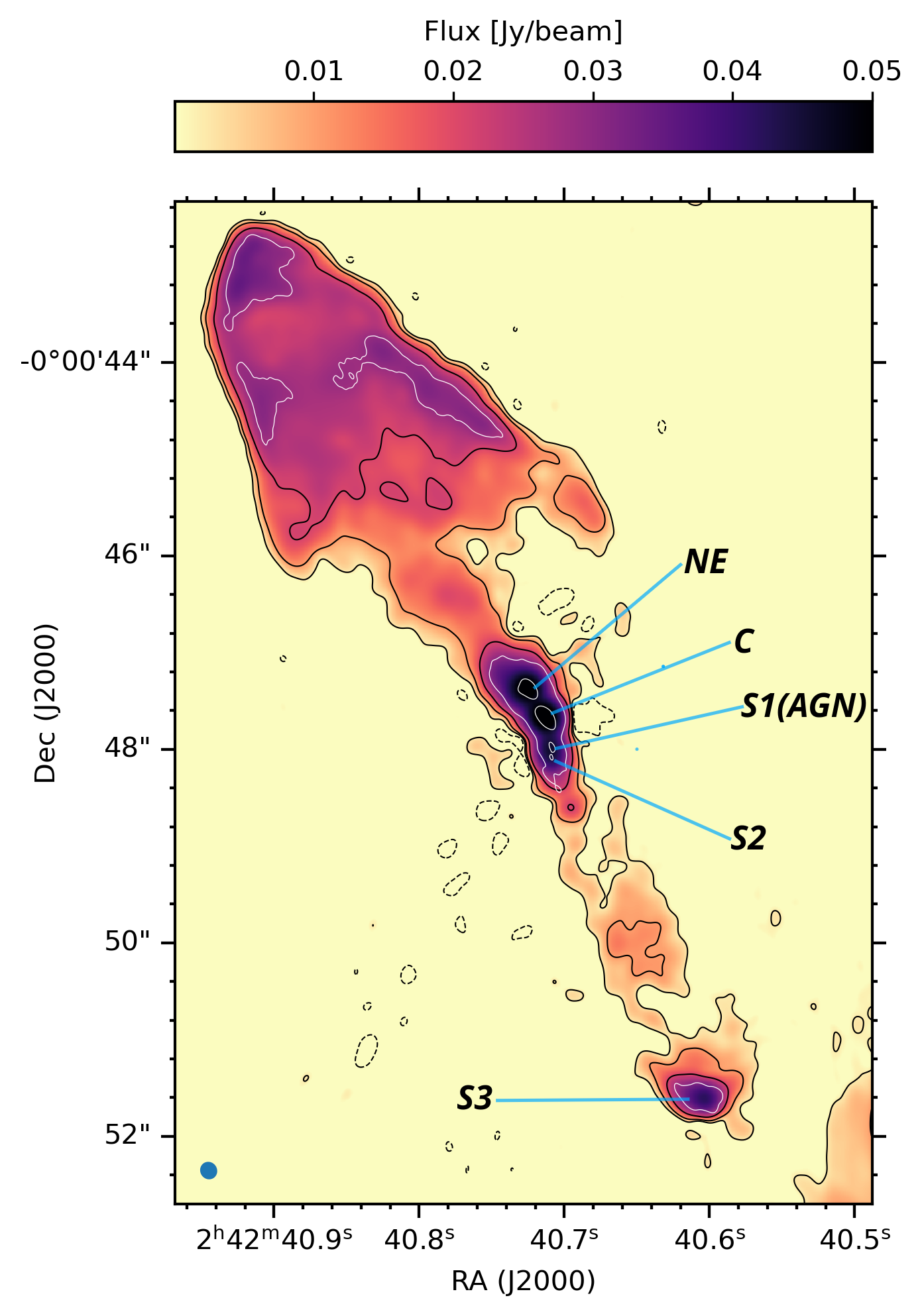}
    \caption{Image of the inner jet in NGC~1068, showing the AGN (S1) and the compact jet components in context, taken from \citet{mutie-2024}. The composite image is constructed from combining $e$-MERLIN+VLA data between 4 and 12 GHz yielding angular resolution scales between 0.18 and 0.05 arcsec with rms sensitivity of 30 $\mu$Jy beam$^{-1}$.}
    \label{fig:combined}
\end{figure}

\subsubsection*{Flux density measurements}\label{measuremeents}

Before extracting flux densities, astrometry checks were done in all data by overlaying the maps and comparing positions of compact sources by measuring the positions of their peak flux densities per beam and assuming the position of S1 as a prior. Flux density measurements of all components were then obtained by using the estimates file in \verb|CASA| task \verb|IMFIT| where the position of S1 was assumed as prior and beam size fixed at 60~mas in all measurements. The flux densities obtained for each observed frequency are presented in Table~\ref{tab:fluxes}, together with the statistical errors. To account for possible systematic uncertainties in the absolute flux calibrations of the different observatories, we added a systematic error of 10\% in quadrature in all flux density errors.

\section{Spectral Energy Distribution Modelling}\label{sec:sed-fitting}

We have a set of continuum flux densities obtained at a consistent angular resolution and $uv$-coverage in the AGN (region S1). In addition, we include data from observations with a poorer angular resolution as hard upper limits to the SED, as they encapsulate more diffuse emission; we leave a 1-$\sigma$ margin to be conservative with the strict limits adopted. We introduce a robust SED fitting scheme to derive physical information from the measured continuum SED. Here we present a summary of the model that has been used in the study of other RQ AGNs \citep{Shablovinskaya-2024}, and leave further details to be presented in del Palacio et al. (in prep.). 

In general, the SED of a RQ galaxy consists of multiple components, including a diffuse population of cosmic-ray electrons, diffuse ionised gas, dust, and possibly a very compact synchrotron source in the core of the AGN dubbed the corona \citep[e.g.,][]{panessa-2019}. The total emission from the model is the sum of these components. In addition, absorption by ionised gas can affect the emission at low frequencies. We parametrise these components as detailed below, adopting a reference frequency of $\nu_0=100$~GHz. To deal with a large parameter space, we fit the SED using the standard sampler for Markov Chain Monte Carlo \texttt{emcee} \citep{emcee} implemented in the Python package \texttt{Bilby} \citep{bilby}. We assume flat priors over a wide range of values for all fitted model parameters. 

\subsection{Synchrotron from diffuse cosmic-ray electrons}
Diffuse cosmic-ray electrons emit an optically thin synchrotron SED with a steep spectral index $\alpha_\mathrm{sy} \leq -0.5$. We write this as
    \begin{equation}
    \centering
        S_\mathrm{sy}(\nu) = {A_\mathrm{sy}} \, \left( \dfrac{\nu}{\nu_0} \right)^{\alpha_\mathrm{sy}} \,, 
    \end{equation}
where $S_\mathrm{sy}(\nu)$ is the synchrotron spectrum at a given frequency ($\nu$), $A_\mathrm{sy}$ is a normalisation constant. These electrons can be related either to a jet or star formation. 

\subsection{Free--free emission from ionised gas} \label{sec:free}
The ionised gas emits FF radiation with an intrinsic spectral index of $\alpha=-0.1$ where it is optically thin, such that
    \begin{equation}
    \centering
        S_\mathrm{ff}(\nu) = {A_\mathrm{ff}} \, \left( \dfrac{\nu}{\nu_0} \right)^{-0.1} \,, 
    \end{equation} 
where $A_\mathrm{ff}$ is a normalisation constant. We note that at low frequencies this emission can drop as the medium becomes optically thick ($\alpha=2$ instead of $-$0.1). However, this effect is unlikely to be relevant considering that the low-frequency  (<~5~GHz) SED can be dominated by synchrotron emission and/or FF absorption. We thus refrain from including this opacity effect in the model to keep the number of free parameters to a minimum.

\subsubsection{Free--free absorption by diffuse gas} \label{sec:FFA}
The SED can be significantly affected at low frequencies due to FF absorption (FFA) by ionised gas. For simplicity, we assume a diffuse, homogeneous medium but the effects of a clumpy medium can be relatively similar though including additional free parameters and are not likely to be constrained without observations at $\nu < 1$~GHz \citep[e.g.][]{Lacki2013, Ramirez-Olivencia2022}. We thus introduce only one free parameter, namely the frequency at which the diffuse gas becomes optically thick, $\nu_\mathrm{diff}$. The absorption-corrected emission is: 
    \begin{equation}
        S_\mathrm{sy,abs}(\nu) = S_\mathrm{sy,int} e^{-\tau_\mathrm{diff}}, 
    \end{equation}
where $S_\mathrm{sy,int}$ is the intrinsic synchrotron spectrum and $\tau_\mathrm{diff}=(\nu/\nu_\mathrm{diff})^{-2.1}$.

\subsection{Synchrotron emission from a compact corona}
The corona is a very compact region close to the accreting SMBH that is filled with extremely hot plasma. \citet{inoue-2014} proposed that if non-thermal electrons coexist in the corona, they would interact with the strong magnetic fields and generate synchrotron emission. This synchrotron radiation would be optically thick (due to synchrotron self-absorption) at radio-cm frequencies, making it extremely challenging to detect, but it would be optically thin in the high-frequency end (mm--sub-mm regime), hence detectable by sensitive instruments such as ALMA.

Recent works analysing the mm continuum emission from a compact core in AGNs have provided strong support for this idea \citep{behar-2015, behar-2018, kawamuro-2022, ricci-2023, Shablovinskaya-2024}. The synchrotron SED of a compact source as a corona is shaped by SSA at frequencies below $\nu_\mathrm{SSA}$ \citep{inoue-2014}. The SED is a power law with a negative spectral index $\alpha \sim -0.5$ to $-1$ at frequencies above $\nu_\mathrm{SSA}$, whereas it has a positive spectral index of $\alpha = 2.5$ below $\nu_\mathrm{SSA}$. The value of $\nu_\mathrm{SSA}$ and the overall corona SED depend on the properties of the corona (magnetic field, size) and the relativistic electrons there. We thus adopt a physical model to compute this SED self-consistently. The first application of this model was already presented in \cite{Shablovinskaya-2024}. In short, the model is based on that of the hybrid corona by \cite{Inoue_2018}, implemented by adapting the code from \cite{margalit-2021} for synchrotron emission by a hybrid plasma. We parametrise the radius of the corona as $R_\mathrm{c} = r_\mathrm{c} R_\mathrm{g}$, where $r_\mathrm{c}$ is an adimensional parameter, $R_\mathrm{g} \propto M_\mathrm{BH}$ is the gravitational radius, and we adopt a SMBH mass of $M_\mathrm{BH} = 1.66\times~10^7$\Msun \citep{gallimore-2024}. To reduce the number of free parameters, we tie the magnetic field strength to the non-thermal electron population by adopting a scaling between the energy density in the magnetic fields ($U_\mathrm{B}$) and in non-thermal electrons ($U_\mathrm{nt,e}$), namely $\eta_B = U_\mathrm{B}/U_\mathrm{nt,e} = 40$. We then allow the parameters $r_\mathrm{c}$ and $\delta = U_\mathrm{nt,e}/U_\mathrm{th,e}$ (fraction of the energy in non-thermal electrons w.r.t. thermal electrons) to vary, while we fix the temperature and Thompson opacity of the corona, $kT = 166$~keV and $\tau_\mathrm{T}=0.25$, respectively, and the spectral index of the non-thermal electron energy distribution, $p=2.7$. We note that the choice of these last parameters has only a small effect on the SED (del Palacio et al., in prep.).

\subsection{Thermal emission from dust}\label{sec:dust}
Dust emission is modelled as a modified-black body spectrum. This is characterised by the frequency $\nu_{\tau_{1}}$ at which the dust opacity becomes equal to unity, and the index $\beta$ of the opacity coefficient $\kappa_\nu \propto \nu^\beta$ (with $\beta \approx 1$--2), at $\nu < \nu_{\tau_{1}}$ the SED is optically thin and has a spectral index $\alpha = 2+\beta$, while at $\nu > \nu_{\tau_{1}}$ it is optically thick and $\alpha = 2$. This can be parametrised as:
    \begin{equation}
        S_\mathrm{d} (\nu) = {A_\mathrm{d}} \, \left( \dfrac{\nu}{\nu_0} \right)^2 \, \left( 1 - e^{-\tau_\mathrm{d}} \right) \,, 
    \end{equation}
where $S_\mathrm{d}$ is the dust spectrum, $A_\mathrm{d}$ is a normalisation constant and $\tau_\mathrm{d}=(\nu/\nu_{\tau_{1}})^\beta$. Given the lack of sufficient sub-mm and far-IR data to fit the dust component, we simply fix $\beta=2$ and $\nu_{\tau_{1}}=800$~GHz, and constrain only the normalisation $A_\mathrm{d}$. We note that the values adopted for $\beta$ and $\nu_{\tau_{1}}$ have no impact on the results obtained with the current dataset. 

\begin{table}
    \caption{Flux densities extracted by fitting 2D Gaussian from component S1. Similar \texttt{CASA} regions were used in all frequencies per component. The 1.4 and 15\,GHz points are obtained from the literature and have a larger beam size hence are used as upper limits (UL) \citep{muxlow-1996}.}
    \label{tab:fluxes}
    \begin{tabular}{lccc}
         \hline
         Frequency (GHz)  & Interferometer  & Flux (mJy) &rms ($\mu$Jy beam$^{-1}$)\\
         \hline
         1.4 (UL)  & $e$-MERLIN & <36.3$\pm$2.8 & 330 \\
         4.8                & $e$-MERLIN &19.4$\pm$0.3& 43\\
         5.3                 &$e$-MERLIN& 19.1$\pm$0.3 & 27\\
         6.3                 &$e$-MERLIN& 12.6$\pm$0.2  &31\\
         6.8                 &$e$-MERLIN&18.4$\pm$0.2    &31\\
         7.3                &$e$-MERLIN & 16.8$\pm$0.2   & 31\\
         9.5                & VLA & 12.3$\pm$0.1    &37 \\
         10.5               & VLA  &12.2$\pm$0.1             &37 \\
         11.5               & VLA&10.2$\pm$0.1   &37 \\
         15 (UL)  & VLA &  <17$\pm$1             & --\\
        18.9               &  VLA&13$\pm$0.1     & 23 \\
         19.4               &  VLA&12.6$\pm$0.1     & 23 \\
         19.9               &  VLA&12.5$\pm$0.1     &37 \\
         21.9               &  VLA&12.1$\pm$0.1     &53 \\       
         22.3               &  VLA&12.5$\pm$0.1   & 54\\
         22.9               &  VLA&12$\pm$0.1    &52 \\
         94                  &ALMA&11$\pm$0.6     &24 \\         
         100                & ALMA&10.2$\pm$0.5   &24 \\         
         241                & ALMA&10.6$\pm$0.6   &24 \\
         256                & ALMA&9.9$\pm$0.1   &24 \\
         345                & ALMA&16.1$\pm$0.3     &87 \\
         357                & ALMA&13.9$\pm$0.1   &87 \\
         477                &ALMA &18.7$\pm$0.5&     170\\
         483                & ALMA&19.6$\pm$0.4     &170 \\
         688                & ALMA&20.0$\pm$2.5     &80 \\
         697                &ALMA&17.0$\pm$1.7 & 80\\
         706                & ALMA&16.5$\pm$1.9     &75 \\
         \hline
    \end{tabular}
    
\end{table}

\section{Results and Discussion}\label{results&discussion}

In this section, we fit different physically motivated models to the data. First, we fit free--free absorption from diffuse gas and a simple power law model to the data whose results (spectral index) show consistency with the thermal free--free spectrum, hence we consider this a thermal free--free model as discussed in Sect.~\ref{sec:model_pl}. The fit shows that some components are missing, especially on the higher end of the spectrum (>~200~GHz). We then add a corona and dust model motivated by works of \cite{Inoue_2020,michiyama-2023} and \cite{GarciaTorus2016}. This model fits well, but the dust component is unconstrained at all as reported in \citep{GarciaTorus2016}, and instead, all the data points seemed to fit the corona bump well as shown in Fig.~\ref{fig:seds_various} top panel. We then add a synchrotron model as presented in Sect.~\ref{sec:model_all} to fit the lower end of the spectrum <~10~GHz as shown in Fig.~\ref{fig:seds_various} middle panel. This fits well but leaves a kink between 10 and 100~GHz. We then add a clumpy absorption model (section \ref{sec:model_clumpy}) to take care of the kink as shown in Fig.~\ref{fig:seds_various} bottom panel.

In Fig.~\ref{fig:seds_various} we present the SED of the AGN of NGC~1068 (S1) and our model fittings under the above mentioned different assumptions. In very simplified terms, the SED is potentially dominated by diffuse synchrotron at frequencies $\lesssim$2~GHz, FF between $\sim$2--200~GHz, the corona between $\sim$200--700~GHz, and most likely dust at higher frequencies (although only an UL is derived with the current dataset). In addition, the impact of FFA might become relevant at frequencies below $\sim$2~GHz. In this section, we provide a more comprehensive discussion of the SED. In the text, we round some numbers and error bars to simplify the discussion and present the corner plot with the complete details of the posteriors of the fit in Appendix~\ref{sec:appendix}.

\subsection{The SED below 200 GHz} \label{sec:models}
 
It is currently challenging to determine whether the SED is significantly absorbed below 4~GHz, as we lack data below $\sim$\,2~GHz with the required high angular resolution and sensitivity to unambiguously detect and separate S1/AGN from the surrounding emission. (Table~\ref{tab:fluxes}). However, the $\sim$0.016$''$ angular resolution VLBA observations at 1.7  and 1.4~GHz by \citet{Roy_ff-ngc1068-1998} and  \citet{Gallimore2004_Parsec}  respectively, did not detect component S1 and placed upper limits of $\sim$\,0.06~mJy. If we include this upper limit, the SED would become strongly absorbed at frequencies $\lesssim$2~GHz, likely due to FFA in a diffuse medium (e.g \cite{Inoue_2020,michiyama-2023}). The VLBA however detects S1 at frequencies above $\sim$\,5~GHz (e.g \cite{Gallimore2004_Parsec,Roy_ff-ngc1068-1998}). Similar conclusions were reached for 3C\,84, for which VLBA observations at 5~GHz failed to detect the counter jet of its \textit{one-sided} jet, but observations at 15 and 22~GHz did detect it, which was interpreted as FFA due to an absorbing screen in front of the counter jet \citep{vermeulen-1994,walker-1994}.  

However, taking into account the lack of sufficient short spacings in these VLBA observations (with the shortest baseline between Los Alamos and Pie Town at 236~km), we can infer that most of the flux is either resolved out and unrecovered in the VLBA observations. Alternatively, there could be strong FFA. The $e$-MERLIN 1.4~GHz data has an angular resolution of $\sim$\,0.19$''$ and does not adequately separate the emission from the AGN from the surrounding components to enable a reliable flux density to be determined. It is a challenge to combine these data with the VLBA data, since $e$-MERLIN's longest baseline (217~km) and VLBA's shortest baseline (236~km) do not overlap. We used \verb|CASA| task \verb|IMFIT| and fixed some parameters such as the position of the peak flux (guided by the $e$-MERLIN 4.5--7.5~GHz map), major and minor axis and the position angle to extract flux densities that we used as upper limits from $e$-MERLIN 1.4~GHz data.

At higher angular resolutions, the source S1 is detected with the VLBA at 5~GHz and 8.4~GHz (measuring 9.1$\pm$0.8~mJy and 5.4$\pm$0.5~mJy respectively), and showing a high brightness temperature $T_\mathrm{B}\approx4\times10^{6}$~K and a relatively flat spectrum $\alpha\approx-0.17$ \citep{Gallimore2004_Parsec}. More recent VLBA observations at 22~GHz with the High Sensitivity Array (HSA) show a compact but structured source with a flux density of $7\pm1$~mJy \citep{gallimore-2024}. This suggests that the emission from S1 on sub-pc scales does not become steeper at higher frequencies. For reference, we show the SED between 1--30~GHz including these VLBA data in Fig.~\ref{fig:sed_s1_vlbi} for comparison with our data. We highlight that our observation with an angular resolution of $\sim~0.06''\approx4$\,pc (which includes the BLR and the dusty torus) whereas VLBA observations, with an angular resolution of $\sim~0.01''\approx1$\,pc,  probe $\lesssim$\,pc scales. It is therefore unclear at this stage whether the radio-cm emission seen on scales $\lesssim~0.01''$ is produced by hot ionised gas associated with the AGN, or non-thermal emission from an unresolved jet that is partially synchrotron self-absorbed. However, we can already conclude from Fig.~\ref{fig:sed_s1_vlbi} that a significant fraction (if not most) of the emission that we detect on scales of $\sim$~0.06$''$ is filtered out on VLBA observations, and thus produced on relatively larger scales.

Below we describe various SED fittings under different assumptions, concluding that the favoured SED is the one shown in Fig.~\ref{fig:seds_various} middle panel. In all cases, we include the diffuse FFA described in Sect.~\ref{sec:FFA}. We also note that the corona and dust components do not change significantly between the models, as they (indirectly) depend only slightly on the SED at $\nu < 200$~GHz.

\subsubsection{Free--free + corona + dust} \label{sec:model_pl}
 
We use the model described in Sect.~\ref{sec:sed-fitting} but assume that there is no diffuse synchrotron emission. We show the fitted SED in the top panel of Fig.~\ref{fig:seds_various}. The model can reproduce adequately most of the SED ($\chi^2_\mathrm{red} = 2.1$), but the residuals of the fit are quite high between 5--8~GHz, suggesting that this model is incomplete.

To rule out other possibilities, we also fit the low-frequency SED using a general, phenomenological power-law component with an arbitrary spectral index between $-2$ and 1. This component could mimic, for example, the synchrotron emission from the compact radio jet (a partially optically thick synchrotron source). The fit yields $\alpha = -0.15 \pm 0.03$, which is consistent with free--free emission ($\alpha = -0.1$), as reported in \cite{muxlow-1996, gallimore_radios_sub_arcsec,Gallimore2004_Parsec}.

\subsubsection{Synchrotron + free--free + corona + dust}  \label{sec:model_all}
The SED of NGC~1068 at $\nu \lesssim 3$~GHz can be dominated by optically thin, diffuse synchrotron emission, although FFA may also be relevant below $\sim$\,2~GHz. 
The SED fit presented in the middle panel of Fig.~\ref{fig:seds_various} is slightly better than without the synchrotron component ($\chi^2_\mathrm{red} = 1.8$), and is, therefore, the preferred one. However, the fit shows a strong degeneracy between the spectral index of the synchrotron emission, which hits the hard limit of $-$2 imposed in the priors, the intensity of the synchrotron emission, and the value of $\nu_\mathrm{diff}$ (Fig.~\ref{fig:posteriors_S1}). We cannot draw strong conclusions about the low-frequency ($<5$~GHz) part of the SED given the lack of data in this range, and the difficulty of spatially separating emission from S1 and adjacent radio emission within the region. The upper limit at 1.4~GHz gives only a loose constraint to these parameters but is insufficient to fully characterise the SED.

\subsubsection{Clumpy absorption + synchrotron + free--free + corona + dust} \label{sec:model_clumpy}
\begin{figure}
    \centering
    \includegraphics[width=\linewidth]{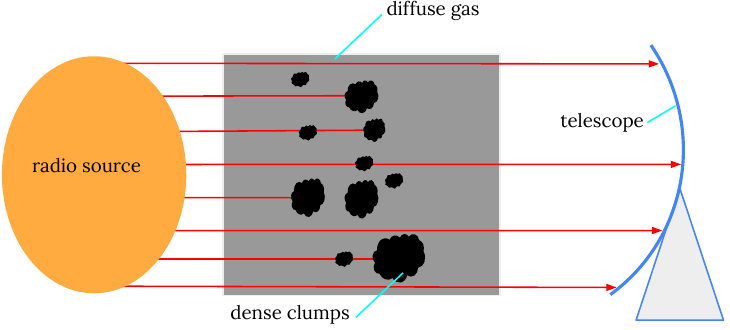}
    \caption{A schematic of the clumpy absorbing medium along the line-of-sight, amidst a diffuse absorbing gas. The clumps become optically thick at a frequency $\nu_\mathrm{cl}$, resulting in absorption of a fraction $f_\mathrm{cov}$ of the total emission (Sect.~\ref{sec:model_clumpy}).}
    \label{fig:schematic}
\end{figure}

There seems to be a small decrease in flux densities between 10--20~GHz that, if real, the previous models cannot reproduce. For this reason, we decided to explore under which conditions a model could account for this behaviour. Following e.g., \cite{Lacki2013}, we assume that the ionised medium is inhomogeneous and clumpy. The FFA absorption depends on the opacity and distribution of the clumps. The clumps can efficiently absorb the synchrotron emission, but also part of the emission propagates between the less dense inter-clump medium, remaining unaffected or with different absorption characteristics \citep[e.g.][]{Lacki2013, Ramirez-Olivencia2022}. A sketch of this scenario is shown in Fig.~\ref{fig:schematic}. This absorption can be modelled in terms of two parameters: the frequency at which the clumps become optically thick, $\nu_\mathrm{cl}$, and the covering factor of the medium, $f_\mathrm{cov}$ \citep[e.g.][for a similar approach]{Conway2018}. In such a case, one can parametrise the absorption-corrected emission as: 
    \begin{equation}
    \centering
        S_\mathrm{sy,abs}(\nu) = S_\mathrm{sy,int} \, \left[ 1 - f_\mathrm{cov} \, (1 - e^{-\tau_\mathrm{cl}}) \right], 
    \end{equation}
where $S_\mathrm{sy,int}$ is the intrinsic synchrotron spectrum and $\tau_\mathrm{cl}=(\nu/\nu_\mathrm{cl})^{-2.1}$. Furthermore, the more diluted gas in-between clumps can contribute with an additional absorption factor $e^{-\tau_\mathrm{diff}}$, with $\nu_\mathrm{diff}\,<\,\nu_\mathrm{cl}$ the frequency at which the diffuse medium becomes optically thick (same as Sect.~\ref{sec:FFA}). We further assume that the free--free emission (Sect.~\ref{sec:free}) also becomes optically thick at $\nu = \nu_\mathrm{cl}$.

\begin{figure}
    \includegraphics[width=0.99\linewidth, trim=0.0cm 1.49cm 0.0cm 0.0cm, clip]{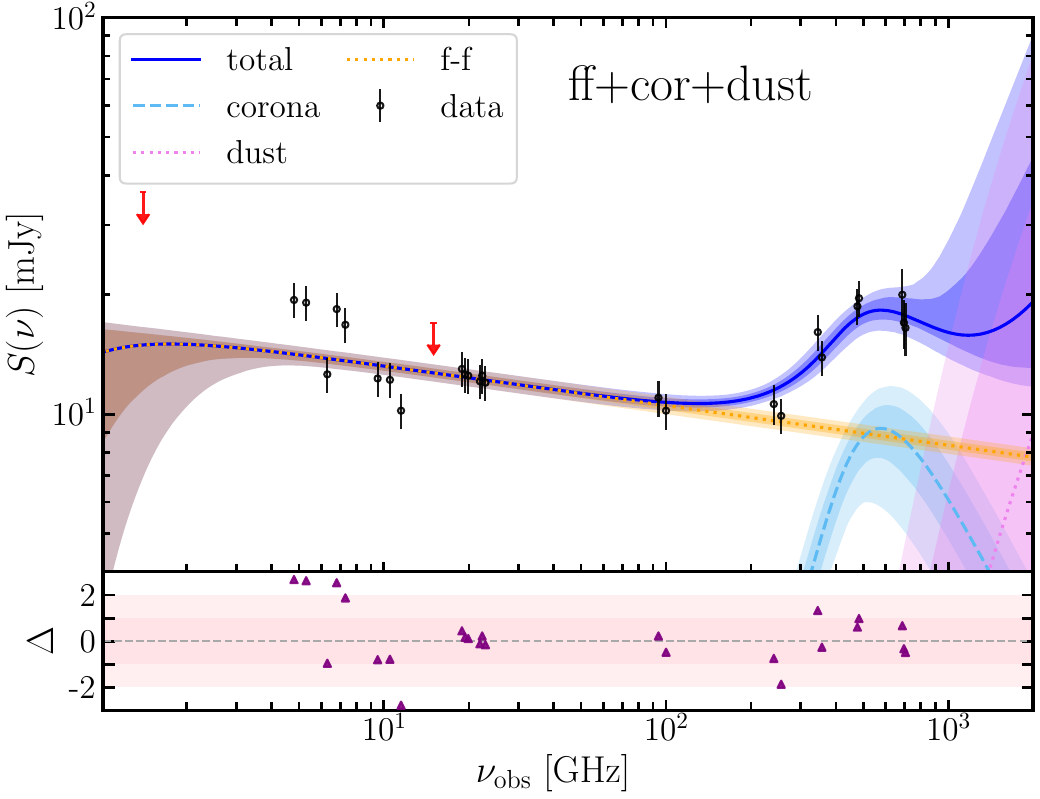}\\
    \includegraphics[width=0.99\linewidth, trim=0.0cm 1.49cm 0.0cm 0.0cm, clip]{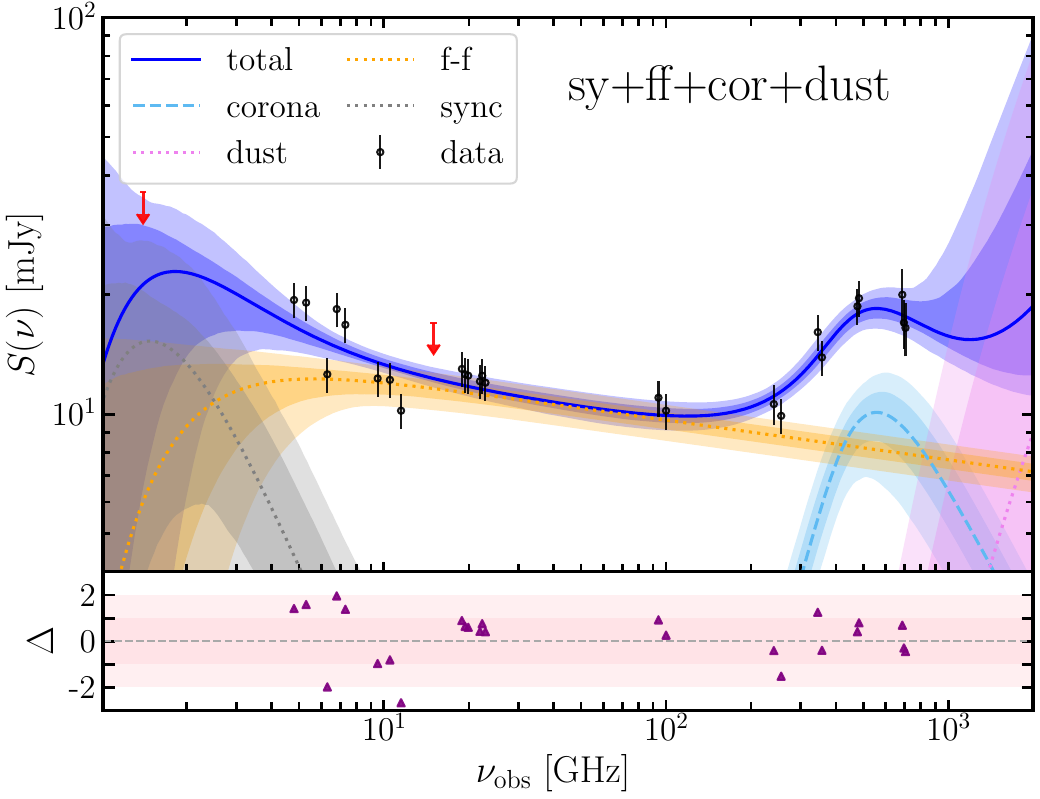}\\
    \includegraphics[width=0.99\linewidth, trim=0.0cm 0cm 0.0cm 0.0cm, clip]{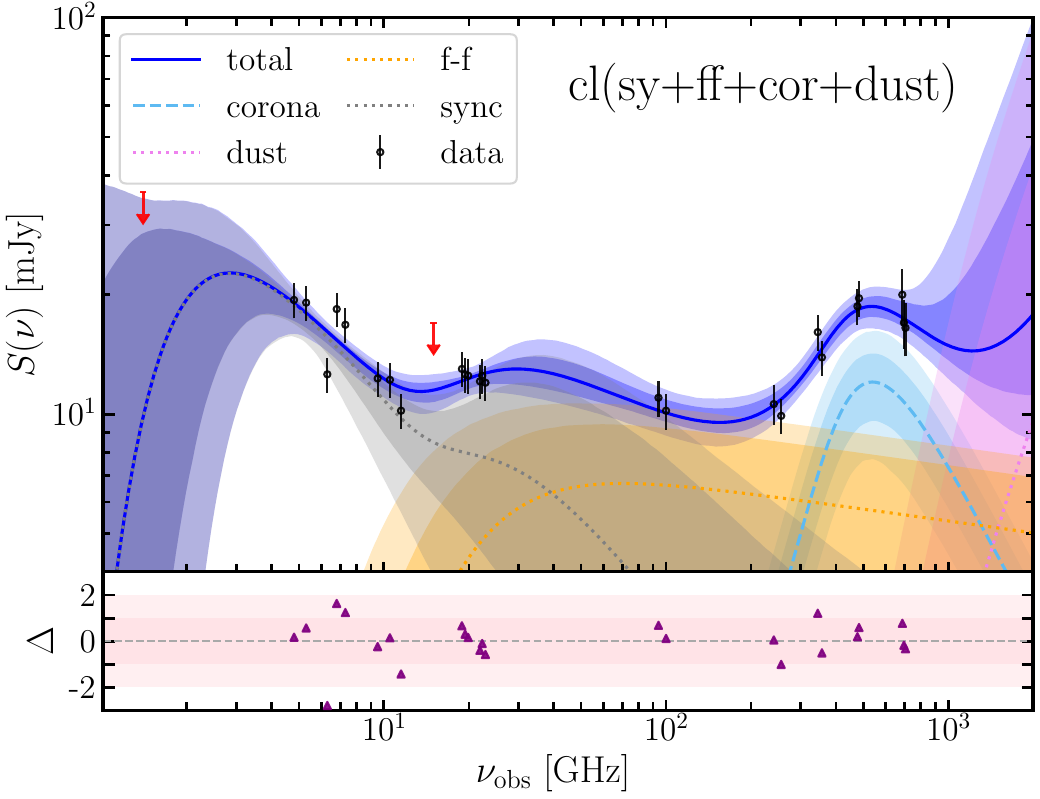} 
    \caption{SED fitting of the S1 source (AGN) in NGC~1068 for the different model assumptions detailed in Sect.~\ref{sec:models}. The shaded regions represent the 1- and 2-$\sigma$ confidence intervals. The bottom sub-panels show the residuals of the fit ( ($S_\mathrm{obs}- S_\mathrm{model}) / S_\mathrm{obs,error}$).}
    \label{fig:seds_various}
\end{figure}

The bottom panel of Fig.~\ref{fig:seds_various} shows the SED fit using this model ($\chi^2_\mathrm{red} = 1.2$). We obtain $\nu_\mathrm{cl} \approx 20$~GHz and $f_\mathrm{cov}\sim 54\%$. Radio sources with turn-over frequencies > 5~GHz have been reported in several works \citep{clemens-2010-hii,o'dea-gps-2021,ballieux-gps-2024}. In \cite{clemens-2010-hii}, they attributed the high turn-over frequencies to \HII\,regions with high densities of ionised gas from rising star formation rates in those sources. In this work, the requirement of $\nu_\mathrm{cl} \sim 20$~GHz would require either even higher densities, beyond those found in the so-called hyper-compact \HII\,regions \citep[][and references therein]{Yang2021}, or larger sizes of the ionised regions. We analyse this in more detail below.

\subsubsection*{High densities?}

We can estimate the density at which $\tau_\mathrm{cl}=1$ from the condition \citep[e.g. using Eq.~1 in][]{Yang2021}:
\begin{equation}
    \left( \frac{v_{\mathrm{cl}}}{\mathrm{GHz}} \right)^{2.1} = 0.082 \, 
    \left( \frac{T_{\mathrm{e}}}{\mathrm{K}} \right)^{-1.35} \, 
    \left( \frac{n_{\mathrm{e}}^2 \, D}{\mathrm{cm}^{-6} \, \mathrm{pc}} \right),
\end{equation}
where $D$ is the characteristic linear size of the ionised region (assuming it is roughly isotropic), and $T_{\mathrm{e}}$ and $n_{\mathrm{e}}$ are the temperature and number density of the electrons, respectively. For simplicity, we fix the electron temperature to the canonical value of 10$^4$~K, and we substitute $v_\mathrm{cl}=20$~GHz. This leads to:
\begin{equation}
\left( \frac{n_\mathrm{e}}{\mathrm{cm}^{-3}} \right)^2 \, 
\left( \frac{D}{\mathrm{pc}} \right) = 1.65 \times 10^{9}.
\label{eq:ne}
\end{equation}
Assuming a characteristic size similar to that of the angular resolution of our observations, i.e. $D \sim 2$~pc, Eq.~\ref{eq:ne} yields a density of $n_\mathrm{e} \sim 3\times 10^{4}~\mathrm{cm}^{-3}$. We note that this corresponds to a column density consistent with the low end of those inferred from spectral modelling of the X-ray SED \citep[$N_\mathrm{H}\,\approx\,10^{23}$--$10^{25}$\cmsq][]{bauer-nustar-2015}. Alternatively, adopting a size similar to BLR clouds, i.e. $D \sim 10^{-5}$~pc, we get a density of $n_\mathrm{e} \sim 10^{7}~\mathrm{cm}^{-3}$, which is about $\sim$1\% of the gas in a typical BLR cloud \citep[e.g.][and references therein]{Netzer2015}. 
We conclude that, in principle, the conditions required for FFA to produce a spectral change between 10--20~GHz seem feasible. However, we cannot rule out other possibilities as well.

\begin{figure}
    \centering
    \includegraphics[width=\linewidth]{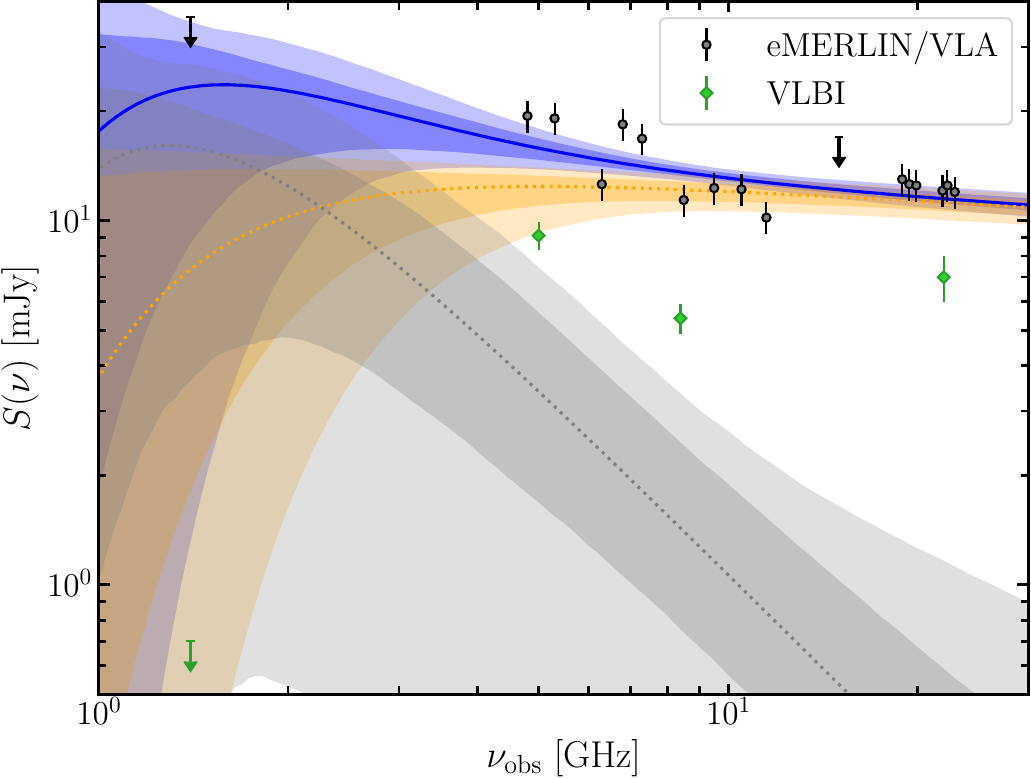}
    \caption{Same as the middle panel of Fig.~\ref{fig:seds_various}, but limited to the radio-cm range. We also show in green diamonds the flux densities from VLBA measurements in the literature. The $e$-MERLIN and VLA data probe consistent angular scales of $\approx$\,60~mas, whereas the VLBA data is only sensitive to $\lesssim$\,10~mas.}
    \label{fig:sed_s1_vlbi}
\end{figure}

\subsubsection*{Is component S1/AGN variable?} \label{sec:variability}

An alternative, plausible explanation for the scatter in the data points between $\sim$\,5--20 GHz could be source variability. Based on causality arguments, variability on time scales of $\sim$\,7 years (the time the 8--12~GHz and 18--23~GHz observations used in this study), implies an emission region of size $\leq 2$~pc. 
This size is consistent with VLBA observations \citep{Gallimore2004_Parsec,Roy_ff-ngc1068-1998,maser_Greenhill_1997} which measure the size of S1 to be $\sim$\,2~pc in diameter. 
The flux densities measured at 5~GHz with the $e$-MERLIN between 1992, 2018 and 2023 (this work) are 12$\pm$2\,mJy, 15.2$\pm$1.2\,mJy, and 19.1$\pm$0.3\,mJy, respectively, while the flux densities measured at $\sim$~20~GHz by the VLA between 1992 and 2015 are 19\,mJy and 12.5$\pm$0.1\,mJy, respectively \citep{mutie-2024}. We note that there may be minor differences between 1992 and later data points due to differing $uv$-coverages of data sets and any flux density scaling difference between the early narrow band VLA and MERLIN and later wideband VLA and $e$-MERLIN observations. However, both 2015 and later data points have been measured using similar procedures \citep[as detailed in][and in this work]{mutie-2024}. The VLA 43 GHz observations conducted between 2000 \citep{jet_cotton_2008} and 2019 (Cotton et al. in prep) reveal a decrease in the flux densities of S1 from 13.1$\pm$0.4~mJy to 10.4$\pm$0.1~mJy, further supporting the variability of S1. Based on these findings, we conclude that source variability is likely to occur in NGC~1068. This variability is consistent with the AGN variability levels of $\pm$10--20\% reported in other local Seyfert galaxies \citep[see][and references therein]{mundell-4151-2003, agn-variability-jones-2011,willliams-ngc4151}.

\subsection{SED between \texorpdfstring{$\sim$}~200 and 706~GHz: The corona}

The flux density of S1 increases between $\sim$\,200 to $\sim$\,500~GHz and flattens (or even decreases slightly) from $\sim$\,500 to $\sim$\,706~GHz. The initial model fits by \cite{Inoue_2020} and \cite{michiyama-2023} favour the presence of synchrotron emission from a compact corona to explain this part of the SED. However, they used data of varied angular resolutions and the effect of it is evident in Fig.~\ref{fig:sed_s1_vlbi} where VLBA fluxes are significantly below $e$-MERLIN and VLA at similar frequencies. The SED shape in the sub-mm rules out that this emission is produced by dust, which has $\alpha > 2$ (Sect.~\ref{sec:dust}); trying to fit an SED without the inclusion of a coronal component leads to a very large $\chi^2_\mathrm{red} = 4.2$ (compared with $\chi^2_\mathrm{red} = 1.8$ with a corona component component fitted as discussed in Sect.~\ref{sec:model_all}) regardless of the values of $\beta$ and $\nu_{\tau_1}$. This result appears to contradict the previous interpretation which assumed a different dust model to explain the sub-mm flux density and was based upon less well-sampled SED data \citep{GarciaTorus2016, michiyama-2023}. 

We adopt flat priors for the corona parameters $r_\mathrm{c}$ and $\log{\delta}$ in the ranges $(20,200)$ and $(-3, 0)$, respectively. Our SED fitting suggests that the corona component dominates the SED between 400--800~GHz, with a peak at $\approx 550$~GHz (Fig.~\ref{fig:seds_various}). The corresponding coronal parameters from the model in Sect.~\ref{sec:model_all} are $r_\mathrm{c}=70\pm 5$ and $\log{\delta}=-1.01\pm0.10$ ($\delta \approx 0.09$), corresponding to $B \approx 148$~G. Depending on the model assumptions (Sect.~\ref{sec:models}), these values can have an additional $\sim$\,10\% dispersion (Appendix~\ref{sec:appendix}). Our results are consistent with those obtained by \citet{Inoue_2020}, considering that they derived $r_\mathrm{c}=20$ assuming $M_\mathrm{BH} = 5\times10^7$\Msun (which would correspond to $r_\mathrm{c} \approx 60$ for the more updated value we adopt of $M_\mathrm{BH} = 1.66\times10^7$\Msun) and fixing $\delta=0.03$. The size of the corona corresponds to $R_\mathrm{c}\approx0.07$~light day (ld), which suggests that its emission can be variable on timescales of $\sim$\,2\,h \citep{Shablovinskaya-2024}. The dust component is very poorly constrained, with essentially only a loose upper limit derived, because of the lack of high angular resolution observations at higher frequencies (in particular, in the far-IR).

We can put our results in the context of the mm and X-ray luminosity correlations found by \cite{kawamuro-2022} and \cite{ricci-2023}. From our SED fitting, we can extract the total mm luminosity at the required frequency and use it to calculate the expected X-ray luminosity; we can then compare this value with that derived from X-ray observations. To accomplish this, we first derived the hard X-ray (unabsorbed) luminosity in the 14--150~keV range using as a reference the intrinsic X-ray luminosity in the 10--40~keV energy range inferred from observations, $L_\mathrm{10-40 \, \mathrm{keV}} \approx 1.5\times 10^{43}~\mathrm{erg}~\mathrm{s}^{-1}$, and the spectral index, $\Gamma= 2.10\pm0.07$ \citep{bauer-nustar-2015}. We obtain $L_\mathrm{14-150\, \mathrm{keV}} = (8.6 \pm 1.3)\times 10^{42}$\ergs assuming a 10\% error on the 10--40~keV luminosity. If we take the luminosity at 230~GHz from our model fitting and use it in the prescription between the 230~GHz luminosity and the 14--150~keV luminosity from \citet{kawamuro-2022} (their table 2 for RQ AGNs), we obtain $L_\mathrm{14-150\, \mathrm{keV}} \approx 2.2 \times 10^{43}$\ergs. Similarly, using the 100\,GHz luminosity from our model and the correlation from Eq.~1 in \citet{ricci-2023}, we obtain $L_\mathrm{14-150\, \mathrm{keV}} \approx 1.9 \times 10^{43}$\ergs. The rough agreement between the inferred values from the correlation and the observations is remarkable considering that our observations at $\sim$\,100--$\sim$~250~GHz will include significant levels of emission not arising from the corona (Fig.~\ref{fig:seds_various}), which might explain why the X-ray luminosity is over-predicted by a factor of $\gtrsim$2. 

\section{Summary and Conclusions}\label{summary}

We used radio--sub-mm data between $\sim 5$--700~GHz, matched in angular resolution at $\sim$~0.06$''$ and $uv$-coverage to probe regions of $\sim$\,2\,pc radius around the SMBH. Our work improves on those from \citet{Inoue_2020} and \citet{michiyama-2023} in that they used data of different angular resolutions. The SED of S1 shows a prominent bump between 200--700~GHz, consistent with synchrotron radiation from a compact corona with radius $R_\mathrm{c} \approx 70\pm5 \,R_\mathrm{g}$, a non-thermal electron population with a fraction of $\sim$\,10\% of the energy density of the thermal electrons, and magnetic field strength of $B \approx 148$~G. The SED of the AGN (S1) is dominated by free--free emission below $\approx$200~GHz, with some poorly constrained levels of synchrotron emission possibly relevant below 3~GHz. There is a hint of a diffuse free--free absorbing medium relevant at frequencies below 2~GHz. The kink in the SED around 10--20~GHz could be either due to intrinsic variability of the source or to a clumpy and high-density absorbing medium.

Future observations at frequencies <\,2~GHz are crucial to further constrain the SED, allowing better characterization of the free--free absorbing diffuse medium and the spectral index of the diffuse synchrotron emission. Such observations are possible with EVN+$e$-MERLIN at 1.4 GHz and with the international LOFAR telescope at $\sim$\,200~MHz, which can provide matched angular resolution and $uv$-coverage to this work. ALMA Band 10 (787--950~GHz) data could also provide important constraints on dust emission. In a follow-up paper, we will investigate in detail the broadband SED of the whole nuclear region of NGC~1068, i.e components NE, C, S2 and S3 (see Fig.~\ref{fig:combined}), providing the conditions of the jet components as well.

\section*{Acknowledgements}

We thank Development in Africa with Radio Astronomy (DARA, Phase 3) for funding this research through the UK's Science and Technologies Facilities Council (STFC) grant ST/Y006100/1. We thank $e$-MERLIN, the VLA, and the ALMA for the observations. $e$-MERLIN is a National Facility operated by the University of Manchester at Jodrell Bank Observatory on behalf of STFC. The VLA and the ALMA are operated by the National Radio Astronomy Observatory (NRAO). The NRAO is a facility of the National Science Foundation operated under a cooperative agreement by Associated Universities, Inc. We acknowledge support from the UK SKA Regional Centre (UKSRC). The UKSRC is a collaboration between the University of Cambridge, University of Edinburgh, Durham University, University of Hertfordshire, University of Manchester, University College London, and the UKRI STFC Scientific Computing at RAL. The UKSRC is supported by funding from the UKRI STFC. SdP and SA acknowledge support from ERC Advanced Grant 789410.


\section*{Data Availability}
The data on which this paper is based are publicly available from the $e$-MERLIN, VLA and ALMA archives under the project IDs described in table \ref{tab:obs}. Calibrated image products are available upon reasonable request to the corresponding author.


\bibliographystyle{mnras}
\bibliography{example}

\newpage
\appendix

\section{Posterior distribution} \label{sec:appendix}

\begin{figure*}
    \centering
    \includegraphics[width=\linewidth]{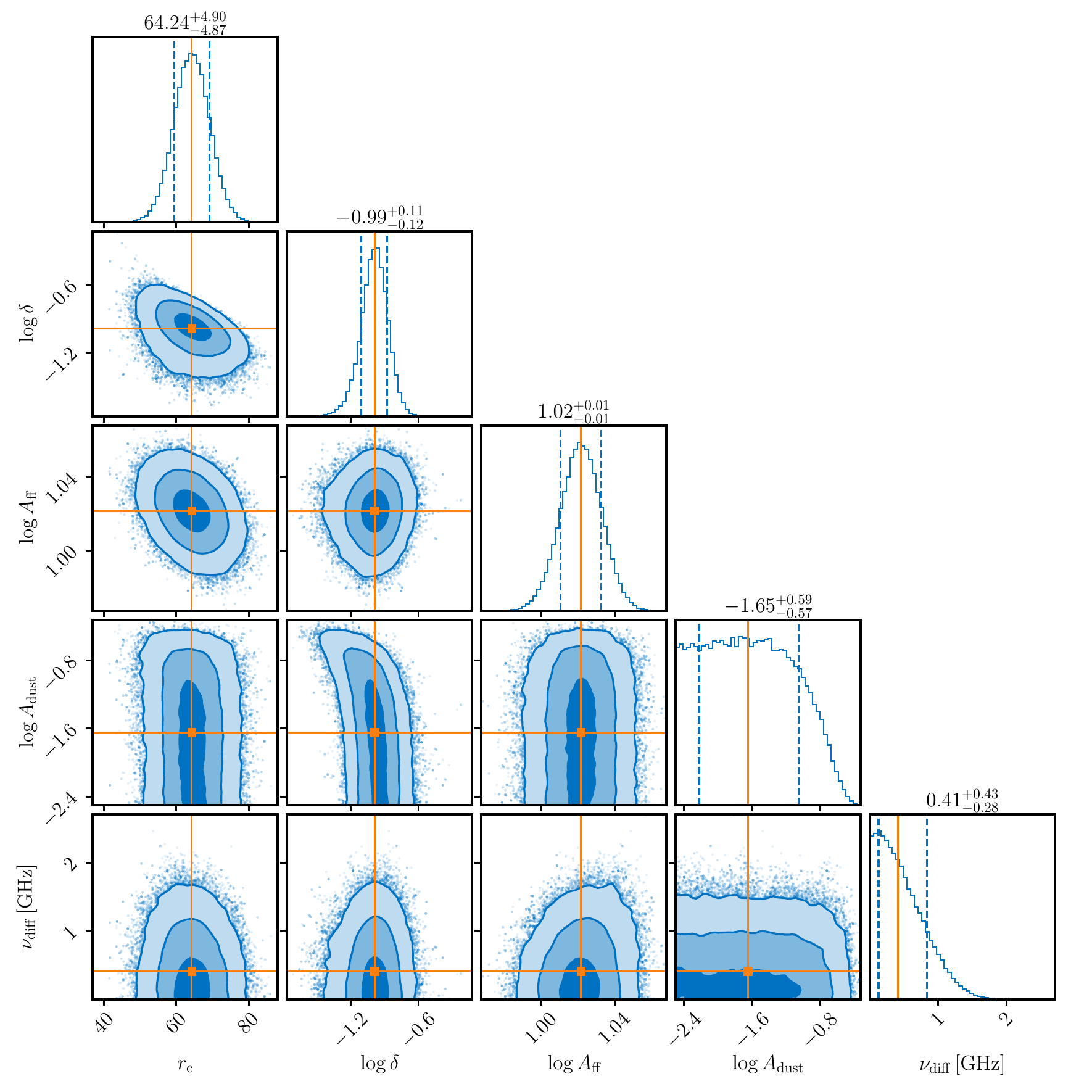}
    \caption{Posteriors of the MCMC fitting in Sect.~\ref{sec:model_pl} (fit shown in Fig.~\ref{fig:seds_various}, top panel). The top panels show the 1-D distributions of each parameter, with the orange line marking the position of the median, and the dashed lines the 1-$\sigma$ confidence interval. We note that for $\log{A_\mathrm{dust}}$ only an upper limit is established.}
    \label{fig:posteriors_fit3}
\end{figure*}

\begin{figure*}
    \centering
    \includegraphics[width=\linewidth]{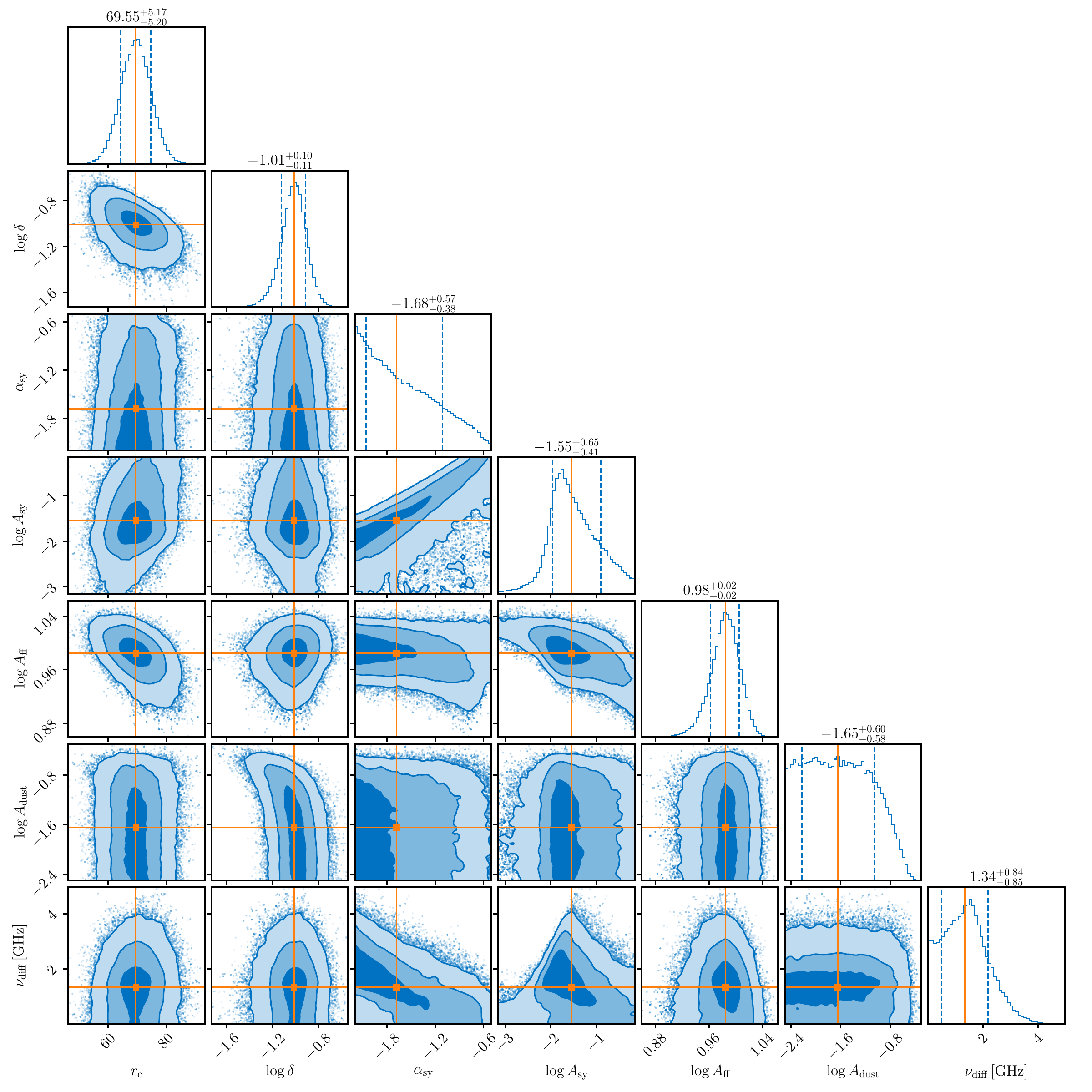}
    \caption{Posteriors of the MCMC fitting in Sect.~\ref{sec:model_all} (fit shown in Fig.~\ref{fig:seds_various}, middle panel). The top panels show the 1-D distributions of each parameter, with the orange line marking the position of the median, and the dashed lines the 1-$\sigma$ confidence interval. We note that for $\alpha_\mathrm{sy}$ and $\log{A_\mathrm{dust}}$ only an upper-limit is established.}
    \label{fig:posteriors_S1}
\end{figure*}

\begin{figure*}
    \centering
    \includegraphics[width=\linewidth]{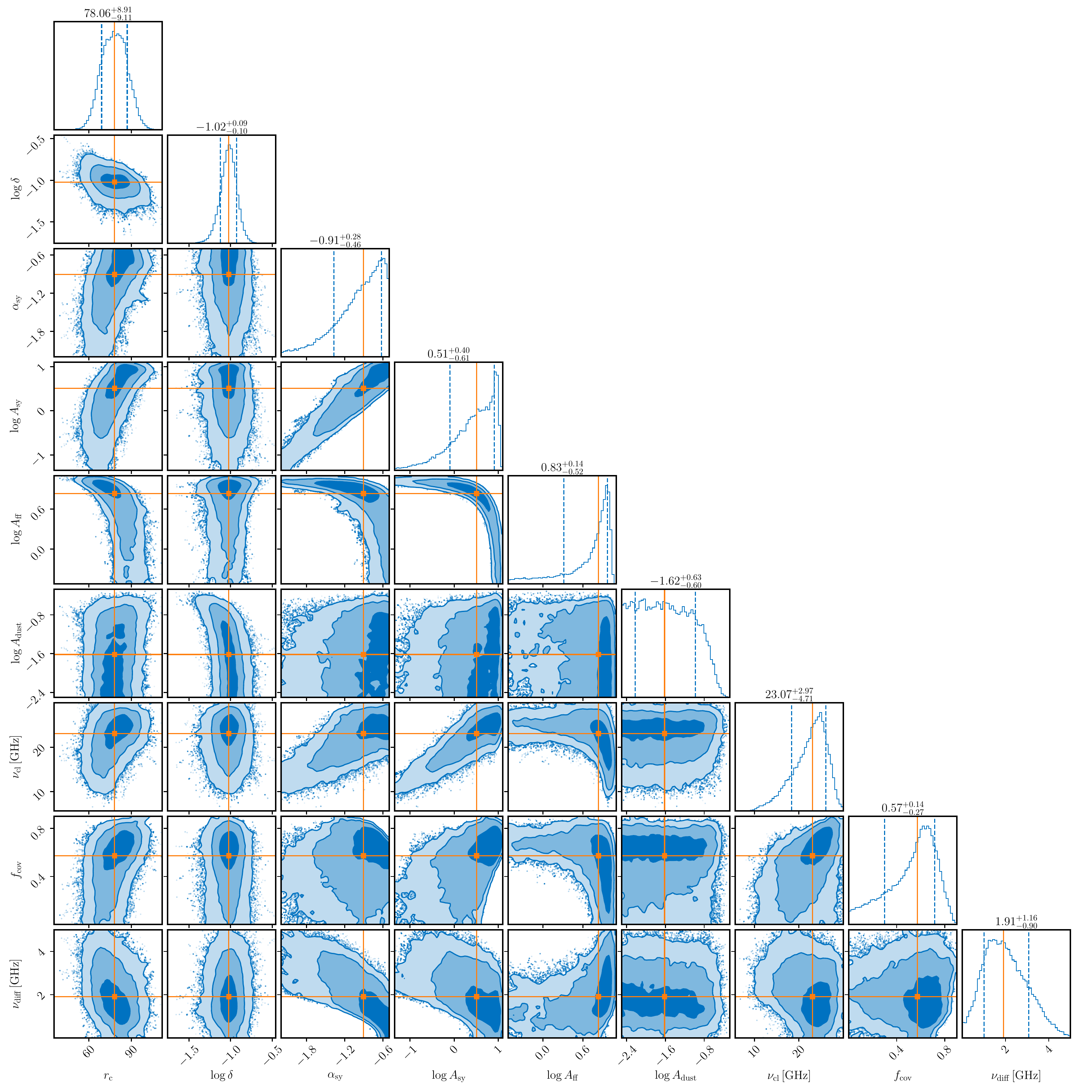}
    \caption{Posteriors of the MCMC fitting using the clumpy model in Sect.~\ref{sec:model_clumpy} (fit shown in Fig.~\ref{fig:seds_various}, bottom panel). The top panels show the 1-D distributions of each parameter, with the orange line marking the position of the median, and the dashed lines the 1-$\sigma$ confidence interval. We note that for $\alpha_\mathrm{sy}$ and $\log{A_\mathrm{dust}}$ only an upper-limit is established.}
    \label{fig:posteriors_clumpy}
\end{figure*}

In Figs.~\ref{fig:posteriors_fit3}--\ref{fig:posteriors_clumpy} we present the posterior distributions of the fitted parameters for S1 in Sect.~\ref{sec:models}. The top panels of each plot show the 1-D distributions of each parameter. The plots are obtained by the MCMC fitting described in Sect.~\ref{sec:sed-fitting}. 

We focus on Fig.~\ref{fig:posteriors_S1}, the cornerplot for the preferred model (Sect.~\ref{sec:model_all}). Despite the broadband coverage of our data, significant degeneracies between the parameters persist, especially for the synchrotron and FFA components, which are relevant at low frequencies where only an upper limit at 1.4~GHz is available. We note that the parameter $\alpha_\mathrm{sy}$ hits the hard limit at $-$2 imposed in the priors, as a steeper synchrotron spectrum is most likely unphysical.

\bsp	
\label{lastpage}
\end{document}